\documentclass[3p,times,procedia]{elsarticle}
\usepackage{nupha_ecrc}
\usepackage{hyperref}

\volume{00}

\firstpage{1}

\journalname{Nuclear Physics A}

\runauth{}

\jid{nupha}

\jnltitlelogo{Nuclear Physics A}

\usepackage{amssymb}

\biboptions{comma,sort&compress}

\usepackage[figuresright]{rotating}

\newcommand{\fm}{\, {\rm fm}}

\newcommand{\GeV}{\, {\rm GeV}}
\usepackage[dvipsnames]{xcolor}

\begin{document}

\begin{frontmatter}



\dochead{XXVIIIth International Conference on Ultrarelativistic Nucleus-Nucleus Collisions\\ (Quark Matter 2019)}

\title{Chemical freeze-out parameters of net-kaons in heavy-ion collisions}


\author[a]{Paolo Alba}
\author[b]{Rene Bellwied}
\author[c]{Valentina Mantovani-Sarti}
\author[d]{Jacquelyn Noronha-Hostler}
\author[b,e]{Paolo Parotto}
\author[b]{Israel Portillo-Vazquez}
\author[b]{Claudia Ratti}
\author[b]{Jamie M. Stafford}

\address[a]{
Lucht Probst Associates GmbH, Grosse Gallusstrasse 9, D-60311 Frankfurt am Main, Germany}
\address[b]{
University of Houston, Houston, TX 77204, USA}
\address[c]{
Technische Universit\"at M\"unchen,
James Franck Strasse 1, 85748 Garching, Germany}
\address[d]{
University of Illinois at Urbana-Champaign, Urbana, IL 61801, USA}
\address[e]{
University of Wuppertal, Wuppertal D-42097, Germany}

\begin{abstract}
We study chemical freeze-out parameters for heavy-ion collisions by performing two different thermal analyses. We analyze results from thermal fits for particle yields, as well as, net-charge fluctuations in order to characterize the chemical freeze-out. The Hadron Resonance Gas (HRG) model is employed for both methods. By separating the light hadrons from the strange hadrons in thermal fits, we study the proposed flavor hierarchy. For the net-charge fluctuations, we calculate the mean-over-variance ratio of the net-kaon fluctuations in the HRG model at the five highest energies of the RHIC Beam Energy Scan (BES) for different particle data lists. We compare these results with recent experimental data from the STAR collaboration in order to extract sets of chemical freeze-out parameters for each list. We focused on particle lists which differ largely in the number of resonant states. By doing so, our analysis determines the effect of the amount of resonances included in the HRG model on the freeze-out conditions. Our findings have potential impact on various other models in the field of relativistic heavy ion collisions.
\end{abstract}


\end{frontmatter}


\section{Introduction}

Ultra-relativistic heavy-ion collisions that create the Quark-Gluon Plasma provide an opportunity to study the phase transition from ordinary hadronic matter to the deconfined phase of matter. We can explore the phase diagram of QCD matter and characterize the strong interaction by studying this transition. The crossover transition at low baryon chemical potential occurs around T $\sim$ 156 MeV, terminating at the proposed critical end point, whereafter a first order phase transition may be seen at higher densities \cite{Aoki:2006br,Bzdak:2019pkr}. It is of interest to study where the freeze-out stages, chemical and kinetic, of heavy-ion collisions lie on the phase diagram. Thermal models can be used to determine the freeze-out parameters, $T_f$, $mu_{B,f}$ (and $V_f$), at chemical freeze-out \cite{Andronic:2017pug,Vovchenko:2019pjl,Alba:2014eba}. Thermal fit analyses rely on minimizing the $\chi^2$ for a fit to experimental data of many particle species \cite{Bellini:2018khg,Adamczyk:2017iwn,Andronic:2017pug}. Results from this method have shown a discrepancy between protons and strange baryons. There have been many efforts to resolve this tension including a flavor hierarchy of hadronization and freeze-out temperatures between light and
strange hadrons \cite{Bellwied:2013cta}, further missing Hagedorn states \cite{Bazavov:2014xya,Noronha-Hostler:2014usa,Noronha-Hostler:2014aia}, final state interactions \cite{Steinheimer:2012rd}, or corrections to the ideal HRG model \cite{Lo:2017lym,Andronic:2018qqt,Aarts:2018glk}. Alternatively, another way to determine the freeze-out parameters is by utilizing net-charge fluctuations \cite{Alba:2014eba,Bluhm:2018aei}. These allow for the determination of freeze-out parameters for separate particle species, and therefore, is well-suited to study the tension between light and strange particles. In these proceedings, freeze-out parameters are studied both by a thermal fitting procedure and by utilizing net-charge fluctuations.

\section{Methodology}

We perform thermal fits for particle yields by means of the Thermal FIST package \cite{Vovchenko:2019pjl}. We utilize the ideal HRG model in the Grand Canonical Ensemble, and focus on data from the STAR collaboration for $\sqrt{s_{\mathrm{NN}}} = 200 \GeV$ only for very central collisions 0-5\% for various particle species \cite{Abelev:2008ab,Adams:2006ke,Adamczyk:2017iwn}.  The only other input for the thermal fit analysis is the choice of particle list and resonance decays that are included. For these proceedings, we will focus on the PDG2016+ list, which has been shown to be in agreement with predictions from Lattice QCD by an analysis of partial pressures \cite{Alba:2017mqu}. We compare how the results from the thermal fits depend on the particle lists, and therefore the number of resonances included in the list, in \cite{Alba:2020jir}.

For the determination of freeze-out parameters from net-charge fluctuations, we also utilize an ideal HRG model in order to calculate the charge susceptibilities as derivatives of the pressure in the grand canonical formalism. We employ the HRG model due to the flexibility and adaptability of parameters to match experimental conditions, including cuts on transverse momentum and rapidity. In this study, we calculate the net-kaon fluctuations as well as the net-proton and net-charge fluctuations. The freeze-out parameters extracted from the combined fit of net-proton and net-electric-charge fluctuations are representative of the light hadron freeze-out parameters, as shown in \cite{Alba:2014eba}. We perform a new study on these fluctuations with the PDG2016+ particle list and compare the results to show the effect of the number of resonances included in the HRG model on the fluctuation analysis. It is important to note that we account for isospin randomization in the investigation of the light hadrons as it was shown to have a significant effect on the net-proton fluctuations \cite{Kitazawa:2011wh,Kitazawa:2012at}. In addition the net-kaon fluctuations are calculated with the same particle list, and we compare the results for the two lists for both sets of freeze-out parameters.

\section{Results}
We present results for both a single and a double freeze-out scenario in order to investigate the proposed flavor hierarchy. In the single freeze-out scenario all particles are fitted simultaneously to produce a single freeze-out temperature. On the other hand, in the case of double freeze-out, we perform two fits, one with the light particles together, namely $\pi^{\pm}$, $p$ $(\bar{p})$, and $K^\pm$, and one with only the strange hadrons which includes all species except pions and protons. Table~\ref{tab:STAR} shows the corresponding fit parameters and $\chi^2_{red}$ for each of the scenarios. We find that the strange freeze-out temperature is higher than the one from the fit with only light particles. In addition, the light temperature is significantly smaller than the single freeze-out temperature. This means that the fit in that case is driven by the strange states, as it was discussed in \cite{Alba:2015iva}.

\begin{figure}
\begin{center}
\includegraphics[width=0.8\textwidth]{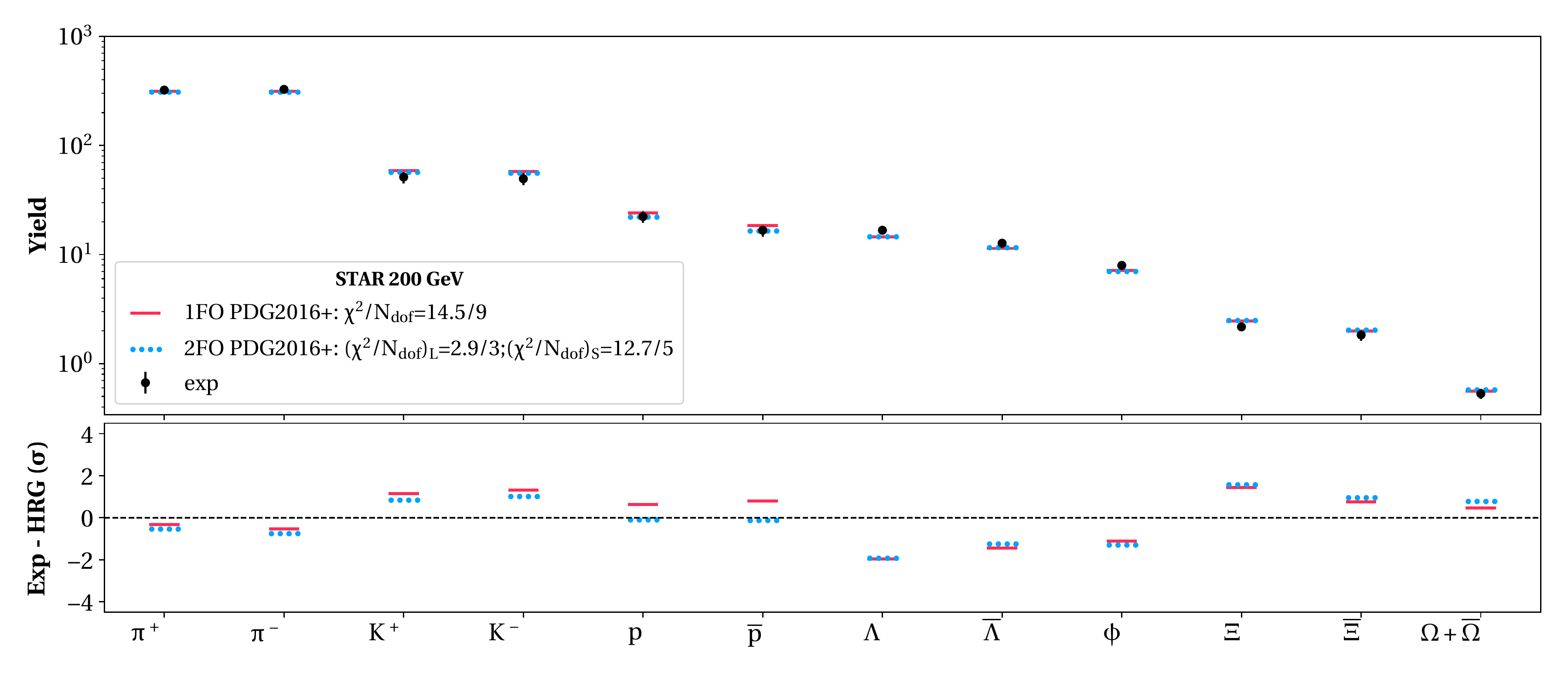}
\caption{
Color online: STAR AuAu $\sqrt{s_{NN}}=200$ GeV data for particle yields in $0-5\%$ central collisions, in comparison to HRG model calculations with the  PDG2016+ resonance list; deviations in units of experimental errors $\sigma$ are 
shown below each panel.
} 
\label{fits}
\end{center}
\end{figure}

The results from the thermal fits are shown in Figure~\ref{fits}. This plot shows the experimental data as black points and the fit results are depicted as horizontal lines, where the solid red line corresponds to the single freeze-out scenario and the blue dotted line shows the results from fitting with two separate sets of freeze-out temperatures and chemical potentials for light and strange particles as defined above. From this plot, we conclude that the double freeze-out scenario has a better overall description of the experimental data, particularly in the light sector. However, the issue with strange baryons still remains and continues to be an open investigation as part of a future work.

\begin{figure}
\begin{center}
\includegraphics[width=0.6\linewidth]{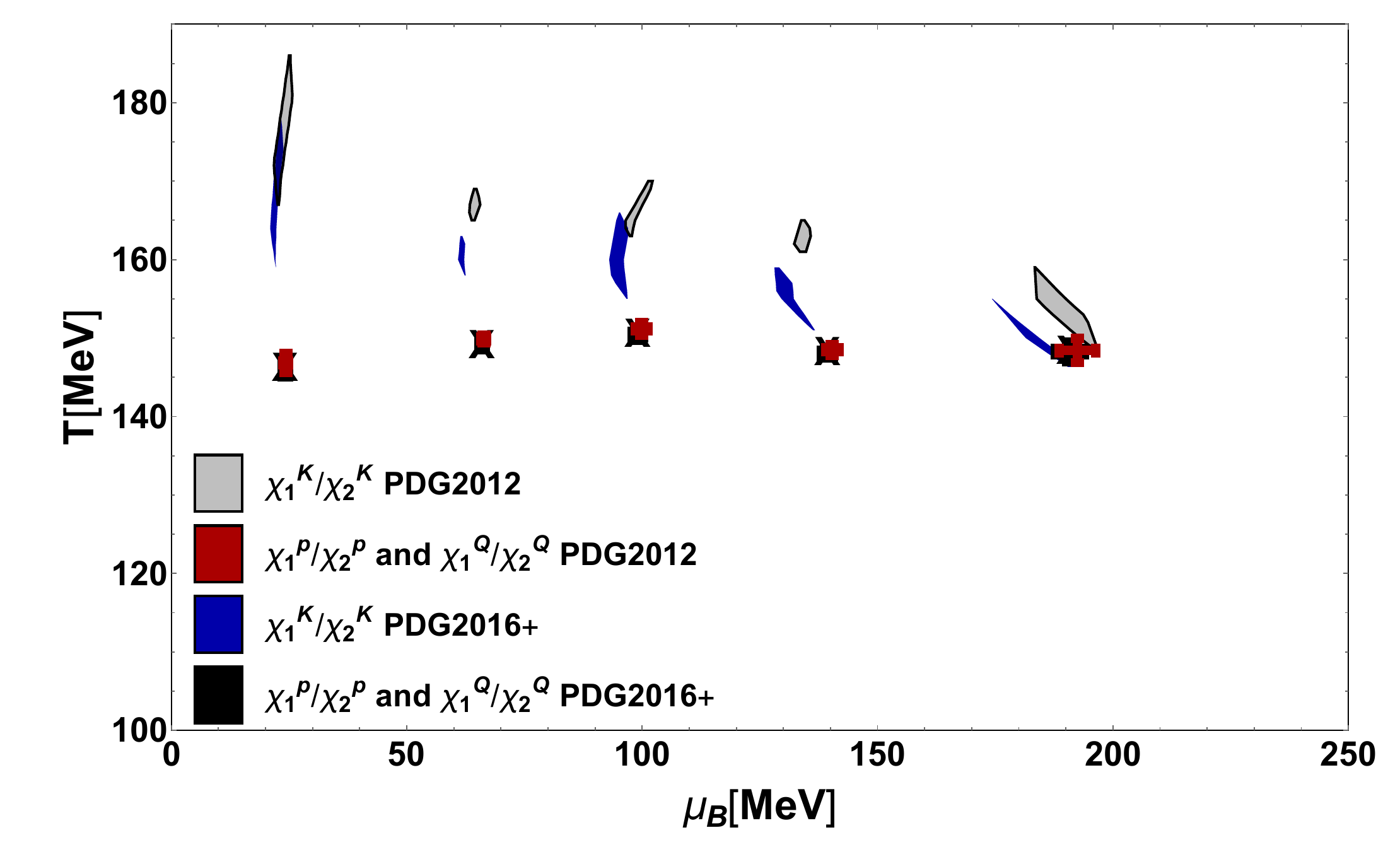}\\
\end{center}
\caption{
(Color online) Comparison of freeze-out parameters obtained via a combined fit of net-proton and net-electric charge with those obtained from the net-kaon fluctuations 
following the analysis from Ref. \cite{Bellwied:2018tkc} with different particle lists.
} 
\label{fig:flucts}
\end{figure}

Figure~\ref{fig:flucts} shows the freeze-out parameters for light hadrons and kaons for the five highest energies of the Beam Energy Scan at RHIC. The light freeze-out conditions are determined by a combined fit of  $\chi_1^p/\chi_2^p$ and $\chi_1^Q/\chi_2^Q$, which are represented by the points in Figure~\ref{fig:flucts}. The red set of points are the results from \cite{Alba:2014eba}, whereas the black points are new calculations with the PDG2016+ particle list for the light hadrons. These new light freeze-out points in turn serve as an important step in the determination of the freeze-out for net-kaons. We use the freeze-out points for the light hadrons to determine the isentropic trajectories through the QCD phase diagram, in order to calculate the net-kaon fluctuations along these trajectories and compare to the experimental data for the net-kaon M/$\sigma^2$, as we have previously shown in \cite{Bellwied:2019pxh}. By doing so, we utilize the isentropes as our additional constraint in the QCD phase diagram such that we can determine the two unknowns, \{$T_{FO}$, $\mu_{B,FO}$\}. The results for net-kaon freeze-out parameters are given by the shapes in Figure~\ref{fig:flucts}: the gray shapes are extracted from \cite{Bellwied:2019pxh} and the blue shapes are those that were calculated by utilizing the PDG2016+ list. This plot shows, firstly, that there is a separation between the light hadron and kaon freeze-out parameters. In addition, the results for the PDG2016+ list as compared to the older hadron resonance list used in the previous analysis, shows that the inclusion of more (strange) resonances is not enough to close the separation between strange and non-strange particles.

\begin{table}
\begin{center}
\begin{tabular}{| c | c | c | c | c | }
\hline
 & T [MeV] & $\mu_B$ [MeV] & Volume [$\fm^3$]  & $\chi^2/d.o.f.$ \\
\hline
Single FO &156.4 $\pm$ 2.2 & 21.1 $\pm$ 6.1 & 2248 $\pm$ 304 & 14.5/9\\
\hline
Light & 152.5 $\pm$  4.9 & $22.4\pm 13.9$ & 2631 $\pm$ 654  & 2.9/3 \\
\hline
Strange & 158.4 $\pm$  3.1  & $20.3 \pm 7.8$ & 1981 $\pm$ 403 & 12.7/5 \\
\hline
\end{tabular}
\caption{Temperature and baryon chemical potential from thermal fits with both single and double
chemical equilibrium temperatures for total yields from STAR 
data~\cite{Abelev:2008ab,Adams:2006ke} for $0-5\%$ centrality in AuAu collisions at 
$\sqrt{s_{NN}} = 200 \GeV$. The last columns show the volume and chi square per number of 
degrees of freedom in the fits.}
\label{tab:STAR}
\end{center}
\end{table}

\section{Conclusions}
We present results on the freeze-out parameters from two different thermal analysis techniques, via fits of particle yields and via net-charge fluctuations. We have found that there is evidence for two separate chemical freeze-out temperatures in heavy-ion collisions for light and strange particles. We see a better description of the experimental data when utilizing the double freeze-out scenario in the thermal fits. In the case of the fluctuations analysis, we see a clear separation between the light and strange freeze-out points in the phase diagram for the highest collision energies at RHIC.

\vspace{-2mm}

\section{Acknowledgments}
The authors gratefully acknowledge many fruitful discussions with Rene Bellwied, Fernando Flor, and Gabrielle Olinger. This material is based upon work supported by the National
Science Foundation under Grant No. PHY-1654219 and by the U.S. Department of Energy, Office of Science, Office of Nuclear Physics, within the framework of the Beam Energy Scan Theory (BEST) Topical Collaboration. We also acknowledge the support from the Center of Advanced Computing and Data Systems at the University of Houston. P.P. also acknowledges support from the DFG grant SFB/TR55. J.N.H. acknowledges the support of the Alfred P. Sloan Foundation, support from the US-DOE Nuclear Science Grant No. de-sc0019175. R.B. acknowledges support from the US DOE Nuclear Physics Grant No. DE-FG02-07ER41521.





\bibliography{all}

\begin{thebibliography}{10}
\expandafter\ifx\csname url\endcsname\relax
  \def\url#1{\texttt{#1}}\fi
\expandafter\ifx\csname urlprefix\endcsname\relax\def\urlprefix{URL }\fi
\expandafter\ifx\csname href\endcsname\relax
  \def\href#1#2{#2} \def\path#1{#1}\fi

\bibitem{Aoki:2006br}
Y.~Aoki, Z.~Fodor, S.~D. Katz, K.~K. Szabo, Phys. Lett. B643 (2006) 46--54.
\newblock \href {http://arxiv.org/abs/hep-lat/0609068}
  {\path{arXiv:hep-lat/0609068}},

\bibitem{Bzdak:2019pkr}
A.~Bzdak, S.~Esumi, V.~Koch, J.~Liao, M.~Stephanov, N.~Xu\href
  {http://arxiv.org/abs/1906.00936} {\path{arXiv:1906.00936}}.

\bibitem{Andronic:2017pug}
A.~Andronic, P.~Braun-Munzinger, K.~Redlich, J.~Stachel, Nature 561~(7723)
  (2018) 321--330.
\newblock \href {http://arxiv.org/abs/1710.09425} {\path{arXiv:1710.09425}},

\bibitem{Vovchenko:2019pjl}
V.~Vovchenko, H.~Stoecker, Comput. Phys. Commun. 244 (2019) 295--310.
\newblock \href {http://arxiv.org/abs/1901.05249} {\path{arXiv:1901.05249}},

\bibitem{Alba:2014eba}
P.~Alba, W.~Alberico, R.~Bellwied, M.~Bluhm, V.~Mantovani~Sarti, M.~Nahrgang,
  C.~Ratti, Phys. Lett. B738 (2014) 305--310.
\newblock \href {http://arxiv.org/abs/1403.4903} {\path{arXiv:1403.4903}},

\bibitem{Bellini:2018khg}
F.~Bellini, Nucl. Phys. A982 (2019) 427--430.
\newblock \href {http://arxiv.org/abs/1808.05823} {\path{arXiv:1808.05823}},

\bibitem{Adamczyk:2017iwn}
L.~Adamczyk, et~al., Phys. Rev. C96~(4) (2017) 044904.
\newblock \href {http://arxiv.org/abs/1701.07065} {\path{arXiv:1701.07065}},

\bibitem{Bellwied:2013cta}
R.~Bellwied, S.~Borsanyi, Z.~Fodor, S.~D. Katz, C.~Ratti, Phys. Rev. Lett. 111
  (2013) 202302.
\newblock \href {http://arxiv.org/abs/1305.6297} {\path{arXiv:1305.6297}},

\bibitem{Bazavov:2014xya}
A.~Bazavov, et~al., Phys. Rev. Lett. 113~(7) (2014) 072001.
\newblock \href {http://arxiv.org/abs/1404.6511} {\path{arXiv:1404.6511}},

\bibitem{Noronha-Hostler:2014usa}
J.~Noronha-Hostler, C.~Greiner\href {http://arxiv.org/abs/1405.7298}
  {\path{arXiv:1405.7298}}.

\bibitem{Noronha-Hostler:2014aia}
J.~Noronha-Hostler, C.~Greiner, Nucl. Phys. A931 (2014) 1108--1113.
\newblock \href {http://arxiv.org/abs/1408.0761} {\path{arXiv:1408.0761}},

\bibitem{Steinheimer:2012rd}
J.~Steinheimer, J.~Aichelin, M.~Bleicher, Phys. Rev. Lett. 110~(4) (2013)
  042501.
\newblock \href {http://arxiv.org/abs/1203.5302} {\path{arXiv:1203.5302}},

\bibitem{Lo:2017lym}
P.~M. Lo, B.~Friman, K.~Redlich, C.~Sasaki, Phys. Lett. B778 (2018) 454--458.
\newblock \href {http://arxiv.org/abs/1710.02711} {\path{arXiv:1710.02711}},

\bibitem{Andronic:2018qqt}
A.~Andronic, P.~Braun-Munzinger, B.~Friman, P.~M. Lo, K.~Redlich, J.~Stachel,
  Phys. Lett. B792 (2019) 304--309.
\newblock \href {http://arxiv.org/abs/1808.03102} {\path{arXiv:1808.03102}},

\bibitem{Aarts:2018glk}
G.~Aarts, C.~Allton, D.~De~Boni, B.~Jäger, Phys. Rev. D99~(7) (2019) 074503.
\newblock \href {http://arxiv.org/abs/1812.07393} {\path{arXiv:1812.07393}},

\bibitem{Bluhm:2018aei}
M.~Bluhm, M.~Nahrgang, Eur. Phys. J. C79~(2) (2019) 155.
\newblock \href {http://arxiv.org/abs/1806.04499} {\path{arXiv:1806.04499}},

\bibitem{Abelev:2008ab}
B.~I. Abelev, et~al., Phys. Rev. C79 (2009) 034909.
\newblock \href {http://arxiv.org/abs/0808.2041} {\path{arXiv:0808.2041}},

\bibitem{Adams:2006ke}
J.~Adams, et~al., Phys. Rev. Lett. 98 (2007) 062301.
\newblock \href {http://arxiv.org/abs/nucl-ex/0606014}
  {\path{arXiv:nucl-ex/0606014}},

\bibitem{Alba:2017mqu}
P.~Alba, et~al., Phys. Rev. D96~(3) (2017) 034517.
\newblock \href {http://arxiv.org/abs/1702.01113} {\path{arXiv:1702.01113}},

\bibitem{Alba:2020jir}
P.~Alba, V.~M. Sarti, J.~Noronha-Hostler, P.~Parotto, I.~Portillo-Vazquez,
  C.~Ratti, J.~M. Stafford\href {http://arxiv.org/abs/2002.12395}
  {\path{arXiv:2002.12395}}.

\bibitem{Kitazawa:2011wh}
M.~Kitazawa, M.~Asakawa, Phys. Rev. C85 (2012) 021901.
\newblock \href {http://arxiv.org/abs/1107.2755} {\path{arXiv:1107.2755}},

\bibitem{Kitazawa:2012at}
M.~Kitazawa, M.~Asakawa, Phys. Rev. C86 (2012) 024904, [Erratum: Phys.
  Rev.C86,069902(2012)].
\newblock \href {http://arxiv.org/abs/1205.3292} {\path{arXiv:1205.3292}},

\bibitem{Alba:2015iva}
P.~Alba, R.~Bellwied, M.~Bluhm, V.~Mantovani~Sarti, M.~Nahrgang, C.~Ratti,
  Phys. Rev. C92~(6) (2015) 064910.
\newblock \href {http://arxiv.org/abs/1504.03262} {\path{arXiv:1504.03262}},

\bibitem{Bellwied:2019pxh}
R.~Bellwied, S.~Bors\'anyi, Z.~Fodor, J.~N. Guenther, J.~Noronha-Hostler,
  P.~Parotto, A.~P\'asztor, C.~Ratti, J.~M. Stafford, Phys. Rev. D 101 (2020)
  034506.
\newblock

\bibitem{Bellwied:2018tkc}
R.~Bellwied, J.~Noronha-Hostler, P.~Parotto, I.~Portillo~Vazquez, C.~Ratti,
  J.~M. Stafford, Phys. Rev. C99~(3) (2019) 034912.
\newblock \href {http://arxiv.org/abs/1805.00088} {\path{arXiv:1805.00088}},

\end{thebibliography}







\end{document}